\documentclass[doublecol]{epl2} 

\usepackage{amsmath} 
\usepackage{amssymb} 
\usepackage{amsfonts}

\usepackage{graphicx}
\usepackage{dcolumn}
\usepackage{bm}
\usepackage{color}
\usepackage{morefloats}
\title{Extending the Newns-Anderson model to allow nanotransport studies through molecules with floppy degrees of freedom}
\shorttitle{Newns-Anderson model in floppy molecules} 
\author{Ioan B\^aldea \thanks{Email: ioan@pci.uni-heidelberg.de. 
Also at 
NILPR, 
ISS, POB MG-23, RO 077125 Bucharest, Romania.}}
\shortauthor{I. B\^aldea}
\institute{                    
  Theoretische Chemie, Physikalisch-Chemisches Institut,
                       Universit\"{a}t Heidelberg, Im Neuenheimer Feld 229,
                       D-69120 Heidelberg, Germany
}
\pacs{73.63.Rt}{Nanoscale contacts}
\pacs{81.07.Nb}{Molecular nanostructures}
\pacs{85.65.+h}{Molecular electronic devices}
\abstract{
The Newns-Anderson model is ubiquitous in studies of the molecular transport in the presence of solvent (outer) reorganization. The present work demonstrates that intramolecular reorganization can also be significant for the transport through molecules with floppy degrees of freedom, for which the Newns-Anderson model can be extended. The expressions of the model parameters deduced from  electronic structure calculations for (4, 4')-bipyridine (44BPY) quantitatively differ from those characteristic for outer reorganization due to strong intramolecular anharmonicities. These expressions can be utilized as input in transport calculations for 44BPY-based molecular junctions of experimental interest 
[Science {\bf 301} (2003) 203, J. Am. Chem. Soc. {\bf 130} (2008) 16045, Nano Lett. {\bf 12} (2012) 354].}
\begin{document}
\maketitle
\section{Introduction}
When an electron subjected to a source-drain bias $V$ is transferred from the negative electrode 
into (say,) the empty LUMO of a neutral molecular system 
M$^0$ embedded in a molecular junction, a transient radical anion 
is created in a first process (M$^{0} \to$M$^{\bullet -}$), 
which is subsequently transferred to the positive electrode 
(M$^{\bullet -} \to$M$^{0}$). 
If these charge transfer processes are fast (strong electrode-molecule coupling), 
they cannot be considered as separate events;
electron transport proceeds 
by coherent tunneling. 
The nuclei are frozen at the optimum molecular geometry $\mathbf{Q}_0$.
This picture is ubiquitously adopted to describe vacuum molecular junctions
within NEGF (nonequilibrium Green's functions) approaches \cite{Datta:05}.

\begin{figure}
$ $\\[3.6ex]
\centerline{\hspace*{-0ex}\includegraphics[width=0.35\textwidth,angle=0]{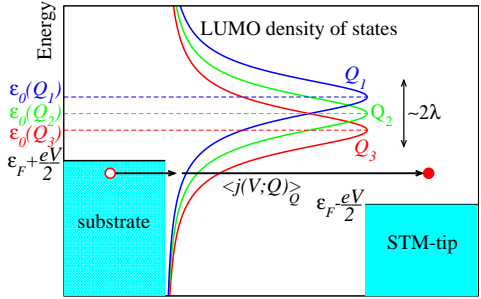}}
$ $\\[0.6ex]
\caption{Schematic representation of the adiabatic molecular transport in the 
Newns-Anderson model [eqs.~(\ref{eq-H-1}) and (\ref{eq-R})]. 
The total current 
represents a $Q$-ensemble average of LUMO-mediated tunneling, whose energy $\varepsilon_0(Q)$ 
fluctuates within an energy range $\sim 2 \lambda$ [cf Eq.~(\ref{eq-R})]]. 
See the main text for details.}
\label{fig:dos}
\end{figure}

For molecular junctions immersed in electrolytes \cite{Wandlowski:08}, 
this picture has a counterpart 
called adiabatic transport 
\cite{Schmickler:93,Zhang:05,Medvedev:07}, which can be summarized as follows.
At a given solvent configuration ${Q}$, 
the molecular device is traversed by a so-called partial electric current $j(V; {Q})$ 
\cite{Schmickler:93,Zhang:05,Medvedev:07}
which is the result of coherent tunneling. 
The slow solvent dynamics with respect to the 
electronic motion 
legitimates the usage of a classical effective coordinate $ {Q}$.
Changes in $Q$ model solvent's dipoles/charges that rearrange 
to stabilize the extra anion's charge and induce significant
variations of the Gibbs adiabatic free energy $\mathcal{G} = \mathcal{G}({Q}; V)$ 
\cite{Schmickler:93,Zhang:05,Medvedev:07}
at a a time scale 
much shorter than the measurement times.
Therefore, as schematically depicted in fig.~\ref{fig:dos}, 
to compare with experiments one has to consider the so-called total current 
$I(V)$, which is obtained by ensemble averaging the partial current with a 
weight function $\mathcal{P}(\mathcal{G}(Q); V)$ that depends on $\mathcal{G}({Q}; V)$ 
[cf.~eq.~(\ref{eq-I}) below].

The Newns-Anderson model \cite{Anderson:61,Newns:69b}
is widely employed in these studies \cite{Schmickler:86,Schmickler:93,Medvedev:07}.
Since early studies on atoms absorbed on a metal surface 
\cite{Newns:69b}, this model continues 
to be used to describe phenomena of recent interest for molecular 
electronics, e.~g., transition voltage spectroscopy  
\cite{Beebe:06,Baldea:2012a,Baldea:2012b,Baldea:2012g}.
It models the molecule as {a} single level $\varepsilon_0$,
which mediates the tunneling between the source and the drain.
This level models the closest 
molecular orbital (LUMO for n-type conduction) to the metal's Fermi energy $\varepsilon_F$.
For a molecule immersed in electrolytes, its energy $\varepsilon_0 = \varepsilon_0({Q})$
fluctuates due to solvent reorganization 
\cite{Schmickler:86,Schmickler:93,Medvedev:07}.

The aim of the present Letter is fourfold: (i) to extend the Newns-Anderson model 
to include intramolecular nuclear reorganization;
to show that this extension (ii) is relevant for molecules of 
current interest in molecular electronics but (iii) the underlying functional dependencies
differ from those describing the solvent reorganization; (iv) to deduce these dependencies 
from electronic structure calculations. {Thus, information will be provided 
that can be subsequently used as input for transport studies based on realistic 
parameters.
Transport calculations and comparison with experiments 
\cite{Tao:03,Venkataraman:06,Venkataraman:12} deserve a special analysis 
and will make the object of a separate publication.}
\section{Specific details}
\label{sec:details}
\begin{figure}
$ $\\[0.6ex]
\centerline{\hspace*{-0ex}\includegraphics[width=0.35\textwidth,angle=0]{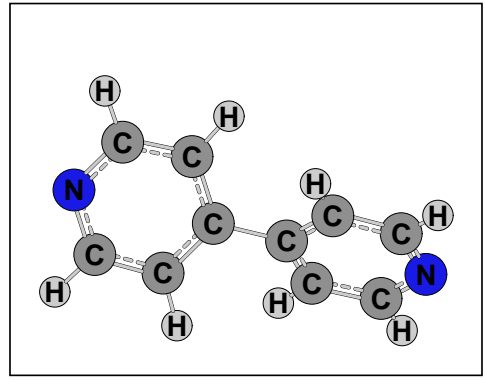}}
$ $\\[0.6ex]
\caption{The neutral 44BPY molecule consists of two pyridine rings 
twisted by $\theta = 37.2^\circ$.}
\label{fig:44bpy}
\end{figure}

As a concrete molecule, (4, 4$^\prime$)-bipyridine (44BPY) will be considered in the present study.
In view of its special structure (cf.~fig.~\ref{fig:44bpy}), 
with two active nitrogen atoms in para positions, 
44BPY is particularly suitable for simultaneous binding to two metallic electrodes.
It has been utilized to demonstrate for the first time the possibility 
of repeated formation of molecular junctions \cite{Tao:03}. 
Compounds based on 44BPY, commonly known as ``viologens'', attracted 
considerable attention for many decades.
44BPY molecules have been incorporated in 
redox active tunneling junctions to demonstrate the 
LUMO electrolyte gating \cite{Wandlowski:08}. Several theoretical 
studies devoted to electron transport 
in 44BPY 
\cite{Hou:05,Thygesen:05c,Bagrets:08,Venkataraman:12} 
considered the coherent tunneling at fixed geometry 
but did not examine the impact of intramolecular reorganization.

The quantum chemical calculations for the present study 
have been done with the Gaussian 09 package \cite{g09} 
at density functional theory (DFT) level by using
the B3LYP 
functional.
Basis sets of double-zeta quality augmented with diffuse
functions (Dunning aug-cc-pVDZ) for the light atoms 
and with relativistic 
core potential (cc-pVDZ-PP from ref.~\cite{Puzzarini:05}) for gold have been employed.
\section{The spinless Newns-Anderson model}
\label{sec:na}
The central assumption of the transport approaches based on the Newns-Anderson model is that 
electric conduction is dominated by a single molecular level. As shown below, this should certainly be 
the case for molecular junctions based on 44BPY \cite{Tao:03,Venkataraman:09b,Venkataraman:12}. 
{From $\Delta$-DFT calculations \cite{Sham:88}, we deduced a HOMO-LUMO gap 
$\Delta_{ } = E_C + E_A - 2 E_N = 8.5$\,eV. This quantity,
expressed in terms of the energies ($E$) of the cation and
anion radicals, and neutral species (subscripts $C$, $A$, and $N$, respectively) 
is the counterpart of the so-called charge gap used in solid state or mesoscopic physics 
(see ref.~\cite{Baldea:2008} and citations therein). Screening effects 
narrow down this gap \cite{Thygesen:09c}, but it certainly exceeds the Kohn-Sham 
HOMO-LUMO gap ($\Delta_{KS} =4.97$\,eV \cite{Hou:05}), which is known to drastically 
underestimate $\Delta$.}
If the electrodes' Fermi level were located midway between HOMO and LUMO 
(a situation wherein the Newns-Anderson model would inherently fail, since both HOMO and LUMO should 
contribute significantly), 
the transmission (Gamow formula)
$T \approx \exp( - 2 d \sqrt{2 m/\hbar} \sqrt{\Delta / 2} )$ 
through an \emph{underestimated} energy barrier of a height 
$\Delta/2={\Delta_{KS}/2 = 2.49}$\,eV 
and a spatial extension 
$d = 2 d_{\mbox{N-Au}} + l_{\mbox{44BPY}} \simeq 2 \times 2.4 + 7.11$\,{\AA}$\simeq 11.9$\,{\AA}
would yield a conductance $G/G_0 = T \approx {10^{-9}}$, 
which is completely at odds with the experimental values
($G/G_0 \sim 10^{-2} - 10^{-3}$ \cite{Tao:03,Venkataraman:12}). 
Here, $G_0 = 2 e^2/\hbar = 77.48\,\mu$S is the conductance quantum.
To conclude, the assumption of a dominant MO appears to be reasonable for 44BPY. Whether the LUMO 
(electron/$n$-type conduction) or the HOMO (hole/$p$-type conduction)
is the dominant MO cannot be determined from transport measurements in two-terminal setups alone.
This issue can be addressed, e.~g., in electrolyte gating 
\cite{Wandlowski:08} or thermopower studies \cite{Venkataraman:12}.
Because they revealed an $n$-type conduction, we will restrict ourselves below to
the case of a LUMO-mediated conduction.

Within the most general Newns-Anderson model \cite{Anderson:61,Newns:69b},
the single MO of energy $\varepsilon_0$
it contains can be empty ($n_{\uparrow, \downarrow} = 0$),
single ($n_{\uparrow} + n_{\downarrow} = 1$),
or doubly ($n_{\uparrow} = n_{\downarrow} = 1$) occupied,
corresponding to the neutral, radical anion, and dianion 
species, respectively.
The second-quantized Hamiltonian has the expression
\begin{equation}
\label{eq-H-2}
H_{ } = \sum_{\sigma=\uparrow,\downarrow}
\varepsilon_{0}\left(\mathbf{Q}\right)
n_{ \sigma}+ U_{ } n_{ \uparrow} n_{\downarrow} + \mathcal{E}_{ph}\left(\mathbf{Q}\right) .
\end{equation}
where $n_{ \sigma} = c^{\dagger}_{ \sigma} c_{ \sigma}$ denote electron number operators.
A Hubbard-type interaction accounts for the Coulomb repulsion $U$ between the two spin directions.
In an STM-setup,
the molecule is coupled to two electrodes (substrate $s$ and tip $t$),
The average molecule-electrode couplings $\tau_{s,t}$ determine a nonvanishing
level width characterized by the broadening functions
$\Gamma_{s,t} \propto \tau_{s,t}^{2}$ \cite{Datta:05}.

The active MO is coupled to classical intramolecular (and, if the case, solvent) 
vibrational modes
$\mathbf{Q}$ that reorganize upon charge transfer. They modulate the MO energy
$ \varepsilon_{0}^{0} \to \varepsilon_{0}\left(\mathbf{Q}\right) $
and store an energy $\mathcal{E}_{ph}\left(\mathbf{Q}\right)$.
This yields a $\mathbf{Q}$-dependence of the total energy
${E}\left(\mathbf{Q}\right) \equiv \langle H \rangle$.
For convenience, the energy at the neutral optimum $\mathbf{Q}_{0}$ will be taken as energy zero,  
$E_N\left(\mathbf{Q}_{0}\right) = 0$.

An important issue in molecular transport is whether the double occupancy of the active MO is significant or not.
Albeit entirely correct only if a single-particle picture (on which the
DFT description relies) is applicable, the analysis
can be done by observing that eq.~(\ref{eq-H-2}) 
can be described in terms of
two single electron states of energies $\varepsilon_{1}(Q) = \varepsilon_{0}\left(\mathbf{Q}\right)$ and
$\varepsilon_{2}\left(\mathbf{Q}\right) = \varepsilon_{0}\left(\mathbf{Q}\right) +
U$ ($\varepsilon_{1} < \varepsilon_{2}$, $U > 0$).
The charge transfer efficiency is determined by the
energy differences $\varepsilon_{1,2} - \varepsilon_F$. Fluctuations in $\varepsilon_{0}\left(\mathbf{Q}\right)$
induced by phonons are of the order of the reorganization energy $\lambda$,  which typically amounts
a few tenths of electron volts \cite{Wandlowski:08}.
So, a doubly occupied LUMO gives a significant contribution only
if $U$ is at most of the order of $\lambda$; otherwise, the state of
energy $\varepsilon_{2}$, too high above the Fermi level, is blocked, and only that of energy
$\varepsilon_{1}$ is relevant.

The Coulomb blockade parameter $U_{ }$ cannot be directly
determined from transport data in a simple manner
\cite{Goldhaber-GordonNature:98,Baldea:2009a,Baldea:2010d}.
Within DFT, $U \to U_{KS}$ obtained from 
energy splitting of the Kohn-Sham LUMO ``orbitals''
is $U_{KS} = 1.54$, $1.52$, and $1.26$\,eV for a 44BPY molecule in vacuum, 
in (aqueous) solution, and in solution with one gold atom attached 
at each of the two nitrogen atoms, respectively.
Similar to the case of the DFT HOMO-LUMO gap, $U_{KS}$
drastically underestimates $U$. 
A substantially higher value is obtained via 
the more adequate method of energy differences based on
eq.~(\ref{eq-H-2}),
$U = E_N + E_D - 2 E_A = 1.92$\,eV 
(instead of $1.26$\,eV),
for the last of the three aforementioned situations (subscript $D$ stands for dianion).
Still, what is really important 
for the present purpose is that 
$U$ 
is much larger than the reorganization energies (see below).
Transfer processes with a doubly occupied LUMO 
are energetically too costly and can be ignored 
in electron transport through 44BPY. So, one can safely employ
a spinless Newns-Anderson model Hamiltonian ($n = c^\dagger c$)
\begin{equation}
\label{eq-H-1}
H_{ } = \varepsilon_{0}\left(\mathbf{Q}\right) \, n_{} + \mathcal{E}_{ph}\left(\mathbf{Q}\right) ,
\end{equation}
as done in existing studies, e.~g., 
refs.~\cite{Schmickler:86,Schmickler:93,Medvedev:07}.

The total energies of the radical anion
$E_{A}(\mathbf{Q})$ and neutral species $E_{N}(\mathbf{Q})$ can be used to
microscopically compute the $\mathbf{Q}$-dependence of the 
parameters entering 
eq.~(\ref{eq-H-1})
\begin{equation}
{\mathcal{E}}_{ph}(\mathbf{Q}) = E_{N}(\mathbf{Q}); \ \varepsilon_{0}(\mathbf{Q}) = E_{A}(\mathbf{Q}) - E_{N}(\mathbf{Q}) .
\label{eq-param-NA}
\end{equation}
Notice that the LUMO energy $\varepsilon_{0}(\mathbf{Q})$ expressed above
is measured with respect to the vacuum.
The reorganization energies of the radical anion
($\lambda_{A}$) and the neutral ($\lambda_{N}$) species are important quantities 
defined by
\begin{equation}
\label{eq-lambda-na}
\lambda_{N} = E_N\left(\mathbf{Q}_{A}\right) - E_N\left(\mathbf{Q}_{0}\right);
\lambda_{A} = E_A\left(\mathbf{Q}_{0}\right) - E_A\left(\mathbf{Q}_{A}\right) ,
\end{equation}
where $\mathbf{Q}_{A}$ denoted the radical anion 
optimum geometry.  

All the phenomenological approaches of molecular transport in electrolytes
based on the spinless Newns-Anderson model of which we are aware
(\emph{e.~g.}, 
Refs.~\cite{Schmickler:86,Schmickler:96b,Zhang:05,Medvedev:07,Wandlowski:08})
assume harmonic $\mathbf{Q}$-dependencies of $E_{N}$ and $E_{A}$, and this yields
($\mathbf{Q} \equiv \{Q_{\nu}\}$; henceforth $\mathbf{Q}_0 \equiv \mathbf{0}$)
\begin{eqnarray}
\label{eq-harm}
\mathcal{E}_{ph}\left(\mathbf{Q}\right) & = & \frac{1}{2}\sum_{\nu}\omega_{\nu} Q_{\nu}^{2} , \\
\label{eq-linear}
\delta \varepsilon_{0}\left(\mathbf{Q}\right) & \equiv &  
\varepsilon_{0}\left(\mathbf{Q}\right) -  \varepsilon_{0}\left(\mathbf{Q} = 0\right) 
= 
- \sum_{\nu}g_{\nu} Q_{\nu} .
\end{eqnarray}
In the cases where eqs.~(\ref{eq-harm}) and (\ref{eq-linear}) apply,
the above reorganization energies are equal 
\begin{equation}
\label{eq-lambda-a-n}
\lambda_{N} = \lambda_{A} = \lambda \equiv \sum_{\nu}  \frac{g^2_{\nu}}{2 \omega_{\nu}}.
\end{equation}

By linearly joining the minima of the radical anion 
and neutral species, one can introduce an effective coordinate 
$ \mathcal{R}$,
$Q_{\nu} = \mathcal{R} g_{\nu} /\omega_{\nu}$ and recast 
Eqs.~(\ref{eq-harm}) and (\ref{eq-linear}) as \cite{Schmickler:96b}
\begin{equation}
\label{eq-R}
\mathcal{E}_{ph}(\mathcal{R}) = \lambda \, \mathcal{R}^2 ; \
\varepsilon_{0}(\mathcal{R}) = 
\varepsilon_{0} - 2 \lambda \, \mathcal{R} .
\end{equation}
The configurations $\mathbf{Q}_0$ and $\mathbf{Q}_A$  
correspond to  $\mathcal{R} = 0$ and $\mathcal{R} = 1$, respectively. 
\section{Extending the Newns-Anderson model to describe intramolecular relaxation}
\label{sec:breakdown}
Most intramolecular modes are fast (frequencies $\omega_{\nu}$
comparable to or higher than $\Gamma_{s,t}$) and should be treated quantum mechanically
\cite{Thoss:09,Thoss:10}. Fingerprints of the associated inelastic tunneling are, e.~g., 
the well-known peaks in the second
derivative $d^2I/dV^2$ at resonant voltage values ($e V = \hbar \omega_{\nu}$).
They negligibly reorganize and therefore are not interesting
in the present context. But if (like in 44BPY, see below) 
slow intramolecular vibrations exist that are sufficiently strongly coupled to the 
LUMO, their reorganization during electron transfer is significant and should be considered.
The problem is to investigate whether a certain molecule possesses such modes and to scrutinize whether 
they can be described by eqs.~(\ref{eq-harm}) and (\ref{eq-linear}).

To this aim, we have performed electronic structure calculations for 44BPY. 
The values of the reorganization energies 
found from eqs.~(\ref{eq-lambda-na})
were found different: $\lambda_N = 0.353$\,eV and $\lambda_A = 0.224$\,eV.
The inequality $\lambda_N \neq \lambda_A$ demonstrates that 
the inner reorganization of the 44BPY molecule
cannot be correctly described by eqs.~(\ref{eq-harm}) and (\ref{eq-linear}).  

The reorganization energies $\lambda_{N,A}$ computed via eqs.~(\ref{eq-lambda-na})
represent ``global'' quantities emerging from electronic structure calculations, 
wherein \emph{all} intramolecular vibrational modes are included 
within a classical picture. To gain further insight, we have split
the reorganization energy into contributions of individual molecular vibrational modes.
Composed of 20 atoms, the nonlinear
44BPY molecule has 54 normal vibrations. In the D$_2$ point group symmetry, 
they are distributed as $14 A + 12 B_1 + 14 B_2 + 14 B_3 $;
$10 A + 9 B_1 + 9 B_2 + 9 B_3 $ are in-plane vibrations
and $4 A + 3 B_1 + 5 B_2 + 5 B_3 $ are out-of-plane vibrations 
\cite{Zhuang:07}.
Our extensive calculations 
confirmed the expectation that significant contributions to the 
reorganization energy arise from the in-plane normal modes with A symmetry.
Out of the ten in-plane modes of A symmetry we have identified two
modes that dominate the inner reorganization. As expected, 
they are related
to the main structural differences between 44BPY$^{0}$ and 44BPY$^{\bullet -}$.

One of these modes (normal coordinate $Q_{46}$) is 
related to the so-called quinoidal distortion \cite{Zhuang:07} 
of the neutral molecule upon electron attachment
(44BPY$^{0} + e^{-}\to$ 44BPY$^{\bullet -}$),
with a shortening of the inter-ring C-C bond and of the C-C bond parallel to it, 
and a lengthening of the C-C bond between them as well as of the C-N bond.
Adiabatic energy curves $E_{N,A}(Q_{46})$ 
reveal a virtually perfect harmonic behavior,
which agree with eqs.~(\ref{eq-harm}) and (\ref{eq-linear})
and yield equal partial reorganization energies
$\lambda_{N}^{(46)} = \lambda_{A}^{(46)}$.
Because its frequency 
(the computed value $\omega_{46} = 1642\,\mbox{cm}^{-1}\simeq 0.2$\,eV
excellently agrees with the strong Raman band observed experimentally 
at 1645\,cm$^{-1}$ \cite{Zhuang:07}) exceeds typical 
$\Gamma_{s,t}$-values \cite{Baldea:2012a,Baldea:2012b,Baldea:2012g},
this mode is too fast to significantly reorganize by the 
classical thermal activation considered in this study.

\begin{figure}
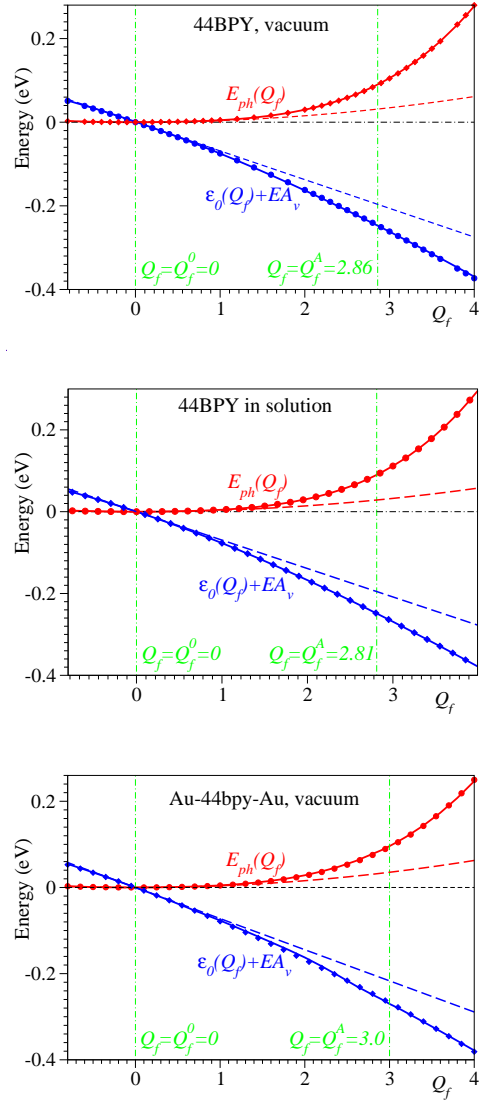

$ $\\[1.0ex]
\centerline{\hspace*{-0ex}\includegraphics[width=0.35\textwidth,angle=0]{Fig3a.eps}}
$ $\\[0.2ex]
\centerline{\hspace*{-0ex}\includegraphics[width=0.35\textwidth,angle=0]{Fig3b.eps}}
$ $\\[0.2ex]
\centerline{\hspace*{-0ex}\includegraphics[width=0.35\textwidth,angle=0]{Fig3c.eps}}
\caption{The model parameters $\varepsilon_{0}$ and $\mathcal{E}_{ph}$ of a 44BPY molecule (a) in vacuum, (b) in 
aqueous solution, and (c) with two gold atoms attached.
The relevant range of the normal coordinate of the floppy degree of freedom $Q_f$
includes values around the minima at $Q_f^{0}$ and $Q_f^{A}$ of the neutral and radical anion species,
respectively. The results of the DFT calculations (solid lines) 
can be excellently fitted with polynomials [eqs.~(\ref{eq-fit}) and table \ref{table}]
(represented by points), but they significantly depart
from the linear and quadratic approximations
of eqs.~(\ref{eq-linear}) and (\ref{eq-harm}), respectively (dashed lines).
See the main text for details.}
\label{fig:e-Q-f}
\end{figure}

The inequality $\lambda_{N} \neq \lambda_{A}$ traces back to the most
appealing feature of the molecular structure, namely the inter-ring twisting
angle $\theta$. 
The mode directly related to the inter-ring torsional motion
represents the floppy (label $f$) degree of freedom of 44BPY.
Adiabatic
energy curves $E_{N,A}(Q_{f})$ along the normal coordinate $Q_{f}$ 
exhibit strong anharmonicities. Its (harmonic) 
frequency is very low ($\omega_f \simeq 62$\,cm$^{-1}$ in the isolated molecule, cf.~table \ref{table}),
but the impact on reorganization is important
because large amplitude oscillations are significant.
The partial reorganization energy of the radical anion $\lambda_{A}^{(f)} \simeq 0.16$\,eV
is almost two times larger than that of the neutral molecule ($\lambda_{N}^{(f)} \simeq 0.09$\,eV). 
From the adiabatic $E_{N,A}$-curves obtained by quantum chemical calculations, 
the functional dependencies of the model parameters $\mathcal{E}_{ph}\left(Q_{f}\right)$ and 
$\varepsilon_{0}\left(Q_{f}\right)$ can be computed from eqs.~(\ref{eq-param-NA}).
The results of these calculations are collected in fig.~\ref{fig:e-Q-f}a. They show significant 
deviations from the linear and quadratic $Q_f$-dependencies of $\varepsilon_{0}\left(Q_{f}\right)$ 
and $\mathcal{E}_{ph}\left(Q_{f}\right)$, respectively. The latter approximations 
are represented as dashed lines in fig.~\ref{fig:e-Q-f}a.
Fig.~\ref{fig:e-Q-f} also shows that fourth-order polynomials
\begin{eqnarray}
\varepsilon_{0}\left(Q_f\right) & = & 
- EA_v
+ \omega_f \left( e_1 Q_f + e_2 Q_f^2 + e_3 Q_f^3 + e_4 Q_f^4\right) , \nonumber \\
\label{eq-fit}
\mathcal{E}_{ph}(Q_f) & = & \omega_f \left(
Q_{f}^{2}/2 + f_3 Q_f^3 + f_4 Q_f^4\right) .
\end{eqnarray}
very accurately fit the DFT-curves. Values of the vertical electron affinity 
$ EA_{v} \equiv - \varepsilon_{0}(0) \equiv E_N(\mathbf{Q}_0) - E_A(\mathbf{Q}_0)$ 
and the other quantities entering  eq.~(\ref{eq-fit}) are given in table \ref{table}.
\begin{center}
\begin{largetable}
\begin{tabular}{|l@{\hspace{1ex}}c@{\hspace{2.ex}}c@{\hspace{2.ex}}c@{\hspace{2.ex}}c@{\hspace{2.ex}}c@{\hspace{2.ex}}c@{\hspace{2.ex}}c@{\hspace{2.ex}}c@{\hspace{2.ex}}|}
\hline
                        & $\omega_f$ &  $f_3$  & $f_4$   & $e_1$   & $e_2$   & $e_3$    & $e_4$   & $EA_v$ \\
\hline
44BPY in vacuum         & 61.9       & 0.0399  & 0.10038 & -9.0236 & -0.9194 & 0.04239 & -0.0007 & 0.444  \\
44BPY in solution       & 58.18      & 0.0634  & 0.1144  & -9.6301 & -1.0751 & 0.04438 & 0.00115 & 2.206  \\
Au-44BPY-Au in vacuum   & 63.2       & -0.0040 & 0.0943  & -9.0110 & -0.2416 & -0.3563 & 0.05714 & 1.586  \\
\hline
\end{tabular}
\caption{
Results for the 44BPY molecule in vacuum, in aqueous solution, and with two gold atoms connected 
to the two nitrogen atoms. The units for the frequency of the floppy mode $\omega_f$ and the 
vertical electron affinities $EA_v$ are cm$^{-1}$ and eV, respectively. The dimensionless  
coefficients $f_{3,4}$ and $e_{1,2,3,4}$ enter the interpolation formulas of eqs.~(\ref{eq-fit}).
}
\label{table}
\end{largetable}
\end{center}
\section{Beyond the case of an isolated molecule}
\label{sec:nonisolated}
The results discussed above refer to an isolated 44BPY molecule in vacuum.
Cases of breakdown of the harmonic approximation 
for low frequency vibrations in isolated molecules are well known
in molecular physics \cite{Tucker:94}.
However, we are not aware of studies pointing out the failure of the 
harmonic approximation for floppy molecules used to fabricate molecular 
junctions, like in the case discussed above. 
For molecular transport, it is also important to
consider the case of a 44BPY molecule immersed in an electrolyte 
and linked to metallic electrodes \cite{Tao:03,Wandlowski:08,Venkataraman:12}.
 
In the present study on 44BPY in aqueous solution (label $sol$), 
the solvent has been considered 
within the polarized continuum model using the integral equation formalism
(keyword SCRF=IEFPCM in Gaussian 09).
The results (fig.~\ref{fig:e-Q-f}b and table \ref{table})
indicate that the most important solvent effect is 
an almost constant shift in the LUMO energy.  
It is related to the strong 
interaction of the anion with the 
water molecules. 
The largest contribution to the vertical electron affinity 
in solution
comes from the anion's solvation free energy $ \Delta G_{A} \equiv 
E_{A}^{sol}(\mathbf{Q}_{A}^{sol}) - E_{A}(\mathbf{Q}_{A}) 
\simeq -2.01$\,eV. Here, $\mathbf{Q}_{A}^{sol}$ denotes the optimized geometry of the 
radical anion in solution.

As a preliminary step in investigating the impact of electrodes, we have also considered
the case of a 44BPY molecule with two gold atoms attached to the nitrogen atoms. 
The corresponding results (fig.~\ref{fig:e-Q-f}c)
reveal an effect qualitatively similar to that of the solvent. 

So, apart from a nearly constant shift of the LUMO energy, for 
$Q_f$-values of interest (cf.~fig.~\ref{fig:e-Q-f}), 
the model parameters $\varepsilon_0(Q_f)$ and $\mathcal{E}_{ph}(Q_f)$ 
vary within ranges, which appear to be little sensitive to the presence 
of solvents or electrodes. 
The physics behind the similarity exhibited by the these numerical results
is the following. The twisting angle $\theta$ of the 44BPY molecule 
is determined by the competition between the
$\pi$-electronic interaction of the pyridyl fragments, which favors
$\pi$-electrons delocalized between coplanar pyridyl rings
($\theta = 0$), and the steric repulsion between the ortho H atoms,
which is diminished by
a twisted conformation ($\theta \neq 0$) \cite{Kassab:96}. 
The latter prevails in the neutral species
44BPY$^{0}$ (empty/oxidized LUMO), which is nonplanar (fig.~\ref{fig:44bpy}).
By adding an extra electron (occupied/reduced LUMO),
the energy gain resulting from $\pi$-electron delocalization between 
the two rings outweighs the steric repulsion, and the radical anion 
(44BPY$^{\bullet -}$) becomes planar.
Therefore, a structural transition from the twisted 
to the planar conformation can only be suppressed if a significant negative charge is transferred 
to the 44BPY unit. This cannot be achieved by immersing in solution, nor even by 
attaching gold atoms. 
In the latter case, the gold atoms acquire a 
small negative charge ($\sim -0.096e$), 
which has a negligible impact on the torsion angle $\theta$.
\section{Summary and outlook}
\label{sec:conclusion}
The present results emphasize the need to consider the inner relaxation 
in junctions based on molecules with floppy degrees 
of freedom; the reorganization energies deduced here ($\lambda \sim 0.1 - 0.2$\,eV)
are comparable with those for outer (solvent) reorganization \cite{Wandlowski:08}.
{To avoid misunderstandings three comments are in order. 
(i) The 
present analysis of the inner reorganization (conformational distortions) 
has implicitly assumed a \emph{given} (e.~g., atop, hollow, bridge)
contact geometry; variations in the binding 
geometry may be larger 
(e.~g., refs.~\citenum{Venkataraman:06,Wandlowski:11}) but they are of a different nature.
(ii) The $Q_f$ dependence of the partial current and conductance discussed here 
refers to a \emph{given} molecule (44BPY); this is qualitatively different, 
e.~g., from the scaling $G \propto \cos^2 \theta$ discussed previously
\cite{Venkataraman:06,Wandlowski:09,Wandlowski:11}, which refers to 
\emph{various}
derivatives of a molecular family (e.~g., biphenyls) wherein the torsion angle 
$\theta$ is varied, e.~g., by a bridging (alkyl) chain. 
(iii) $\theta$ 
can\emph{not} be specified in terms of the single normal coordinate
$Q_f$ only. Thermal averaging over $\theta$ 
(as done for biphenils \cite{Venkataraman:06})
amounts to consider that, 
\emph{concomitant} with the floppy mode $\omega_f$,
other modes of much higher frequency are also thermally activated.
Because such fluctuations are energetically costly in biphenyls, 
their effect is reduced \cite{Venkataraman:06}.}

While revealing that a spinless Newns-Anderson is justified for 44BPY,
the present study has shown that the ${Q}$-dependencies 
of eqs.~(\ref{eq-harm}) and (\ref{eq-linear}) assumed to work in electrolytes are 
inappropriate for the reorganization of low frequency intramolecular vibrations,
which are characterized by a pronounced anharmonic behavior.

New formulas for $\varepsilon_{0}({Q_f})$
and $\mathcal{E}_{ph}\left({Q_f}\right)$ 
have been deduced from microscopic calculations [cf.~eqs.~(\ref{eq-fit})],
which replace the expressions of eqs.~(\ref{eq-harm}) and (\ref{eq-linear}) used for solvent
reorganization. 
The $Q_f$-dependence of the model parameters is important because the ensemble average needed to 
compute the experimentally relevant quantity, namely the total current $I(V)$, requires an integration 
over $Q_f$
\begin{equation}
\label{eq-I}
I(V) = \langle j\left(V; Q_f\right)\rangle = \int j\left(V; Q_f\right) 
\mathcal{P}\left(Q_f; V\right)
d\,Q_f  .
\end{equation}
The fact that 44BPY possesses a single low frequency vibrational mode that significantly reorganizes
represents an enormous simplification; otherwise,
an ensemble average involving a $\mathbf{Q}$-integration over all of the 
54 internal degrees of freedom would be a formidable numerical challenge.  

The thermal weight function 
$\mathcal{P}(Q_f; V) \propto 
\exp\left[ - \mathcal{G}(Q_f; V)/(k_B T)\right] $
is determined by the Gibbs free energy $\mathcal{G}( Q_f; V)$ of the 
(partial) oxidation state of the molecule 
(i.~e., LUMO occupancy $0 < n(Q_f; V) < 1$) \emph{linked} 
to biased electrodes. $\mathcal{G}$ depends on $Q_f$, $V$,
and (if applicable) on electrolyte's overpotential 
\cite{Wandlowski:08,Schmickler:93,Zhang:05,Medvedev:07}. 
The need to carry out an ensemble averaging is an essential aspect, 
which renders purely ab initio approaches (as used for rigid molecules in vacuum) inapplicable;
therefore, feasible approaches to date have to resort to models. 

The above results can (and will) be used in subsequent studies to deduce the current 
within the adiabatic transport approach, for which eq.~(\ref{eq-I}) can serve as starting point.
Expressions for the partial current $j(V; \varepsilon_{0}(Q_f))$ 
and the level (LUMO) occupancy $n(\varepsilon_{0}(Q_f); V)$ are known 
\cite{Schmickler:86,Medvedev:07,Baldea:2010e,baldea:arXiv1108.1039B}. 
Along with the expression of $\mathcal{E}_{ph}(Q_f)$,
the LUMO occupancy is required to express the 
$Q_f$- (and $V$-)dependent Gibbs adiabatic free energy $\mathcal{G}$.
One should emphasize at this point 
that  $\mathcal{G}(Q_f; V)$ is not simply related to
the adiabatic energies $E_{N,A}(Q_f)$; the former quantity characterizes 
a molecule linked to biased electrodes, while the latter quantities pertain to an isolated molecule.

Earlier studies demonstrated that even a single reorganizable harmonic mode, 
eq.~(\ref{eq-R}), yields highly nontrivial 
adiabatic $\mathcal{G}$-surfaces \cite{Schmickler:86,Medvedev:07}. In view of the
significant anharmonicities
embodied in the expressions adequate for
a floppy degree of
freedom, one can expect more complex adiabatic surface topologies, 
with the need to distinguish between several important limiting cases.
To simply motivate this, one should note that,
unlike in electrolytes,
reorganization effects in floppy molecules cannot be merely characterized with the aid of a single quantity
($\lambda = \lambda_N = \lambda_A$). 

As shown recently \cite{Baldea:2012b,Baldea:2012g}, the impact of fluctuations in the MO-energy offset 
$\overline{\varepsilon}_0 \equiv \varepsilon_0 - \varepsilon_F$
on the ohmic conductance can be significant.
Molecule-electrode interactions can be important sources of such fluctuations \cite{Baldea:2012b}.
In view of the present study, 
in molecular junctions like those based on 44BPY fabricated experimentally 
\cite{Tao:03,Wandlowski:08,Venkataraman:12},
one can expect that, even in the absence of other effects, 
reorganization effects represent an important source 
for large $\overline{\varepsilon}_0$-fluctuations 
($\delta \overline{\varepsilon}_0 \sim 0.2 - 0.4$\,eV).
As a straightforward application, one can employ eq.~(\ref{eq-I}) to compute
conductance histograms, which can be directly compared with those
available from experimental studies \cite{Tao:03,Venkataraman:12}.
\acknowledgments
The author thanks Horst K\"oppel for valuable discussions
and the Deu\-tsche
For\-schungs\-ge\-mein\-schaft (DFG) for financial support.
\end{document}